# High school students' representations and understandings of electric fields

Ying Cao[1,*] and Bárbara M. Brizuela[2]

[1]*Chemical, Biological and Environmental Engineering, Oregon State University, Corvallis, Oregon 97331-2702, USA*
[2]*Education, Tufts University, Medford, Massachusetts 02155, USA*

This study investigates the representations and understandings of electric fields expressed by Chinese high school students 15 to 16 years old who have not received high school level physics instruction. The physics education research literature has reported students' conceptions of electric fields postinstruction as indicated by students' performance on textbook-style questions. It has, however, inadequately captured student ideas expressed in other situations yet informative to educational research. In this study, we explore students' ideas of electric fields preinstruction as shown by students' representations produced in open-ended activities. 92 participant students completed a worksheet that involved drawing comic strips about electric charges as characters of a cartoon series. Three students who had spontaneously produced arrow diagrams were interviewed individually after class. We identified nine ideas related to electric fields that these three students spontaneously leveraged in the comic strip activity. In this paper, we describe in detail each idea and its situated context. As most research in the literature has understood students as having relatively fixed conceptions and mostly identified divergences in those conceptions from canonical targets, this study shows students' reasoning to be more variable in particular moments, and that variability includes common sense resources that can be productive for learning about electric fields.

## I. INTRODUCTION

The concept of an electric field is usually introduced in high school and college physics. Studies that have investigated students' understandings of electric fields have documented ideas prevalently held by students: for example, big charges exert stronger forces [1], field lines represent the trajectories of moving charges [2,3], and electric forces act at a distance with no necessary medium [2].

The research methods primarily adopted in previous studies to assess students' ideas were having students answer textbook-style questions in questionnaires and interviews [1–4]. Because the questions were written in disciplinary-specific language (physics terms, formulas, and principles), such research was insufficient to capture the ideas students may also have but tend to describe in nondisciplinary language. Also, when answering questions written in formal physics vocabulary, students may not interpret the vocabulary in the same way as researchers do [1]. In such cases, we may have misinterpreted student thinking.

In the present study, we investigate students' ideas of electric fields through a different means: tapping into the ideas students express in everyday language and in the representations they produce prior to receiving formal instruction [5,6]. Student-produced representations and the elicited ideas can help researchers better understand student thinking and will be resources for students' future learning [7,8].

The question guiding this research is what ideas of electric fields do high school students have as they are engaged in open-ended tasks before receiving formal instruction?

## II. RESEARCH ON STUDENTS' UNDERSTANDINGS OF ELECTRIC FIELDS

Since the early 1990s, physics education researchers have been studying various aspects of students' understanding of electric fields. The main effort has been to identify students' learning difficulties and speculate on possible reasons for those difficulties. In this section we will review the literature on students' understandings about the concept of electric fields. In our review, we will point out the research methods used in previous studies, given that our method was distinct. When we describe the results of previous studies, we will summarize their main findings and also highlight the way in which their claims relate to our findings.

Galilli [9] examined high school students' conception of electromagnetism by administering a paper-and-pencil test. He found that students did not perform well on the test and argued that it was because students learn mechanics before electromagnetism and the two are centered around very different concepts. According to Galilli [9], when students learn mechanics, force is the key concept and the concept of a field is hardly mentioned. However, when students learn electromagnetism later on, field becomes the central concept. In mechanics, objects make direct contact to exert and experience forces. In electromagnetism, however, a field is the necessary medium for the interaction between electrically charged objects. Galilli argued that students learning electromagnetism might have misinterpreted the nature of electric forces and other related concepts (e.g., work-energy conservation) in electromagnetism because these students had previously learned to interpret these concepts differently in mechanics.

[*]Corresponding author.
caoyin@oregonstate.edu

Eylon and Ganiel [10] and later Thacker, Ganiel, and Boys [11] conducted research studies with university students about their understanding of electricity. The results suggest a gap between students' understanding of electric circuits and that of electrostatics. This gap, the authors proposed, might result from the disconnected conceptions students had in different situations that electric phenomena were described: the macro and micro levels. The former involves numerous electric charges working collectively (such as in electric circuits), while the latter mainly deals with interactions among a handful of individual electric charges (such as in electrostatics).

Viennot and Rainson [4] interpreted Eylon and Ganiel's findings [10] as "amply show[ing] students' difficulties in seeing the role of electric fields in the interplay of the different elements of a circuit" (p. 475), which motivated them to explore students' reasoning about electric fields. They decided to focus on students' comprehension of the principle of superposition: whether students applied the principle of superposition properly and, if not, what were the common obstacles that prevented students from appropriately using the principle. Viennot and Rainson [4] administered a questionnaire that required students' written explanations of physical theories of electric fields. When discussing the findings from their study on students' responses, Viennot and Rainson claimed that they confirmed one of the findings of Rozier and Viennot's study [12] about students' reasoning in thermodynamics: students used linear causal reasoning to solve a problem involving the simultaneous influence of several factors.

Viennot and Rainson [4] found that students in their study generated storylike comments on the physical phenomenon presented to them. They then argued that this finding indicated that the students conceived the electrostatic phenomenon in question as a successive sequence of events; they illustrated the storylike comments with an example, "A charge is placed somewhere, then the insulator gets polarized, and then there is a field created at point M" (p. 479). However, without further information, this particular quote can just state an analytical causal chain without necessarily referring to the events temporally one after another, in which case the student comment was totally legitimate. Also, none of these studies have explicitly addressed whether students understood (or did not understand) that an electric charge experiences forces from multiple other electric charges simultaneously, which is the basic idea of the principle of superposition.

The studies mentioned above related students' ideas of electric fields to what those students had learned about different topics in physics. Allain and Beichner [13], on the other hand, examined students' ideas of electric fields in relation to the concept of rate of change in mathematics. Allain and Beichner studied college students' conception of electric potentials and speculated that students' performance on physics questions of field and potential might correlate to students' performance on mathematics questions of rate of change, because the strength of an electric field is the spatial rate of change, or gradient, of the magnitude of the electric potential of the field. Allain and Beichner developed a questionnaire that contained both types of questions. Analysis of student responses to the questions indicated a pattern: students who answered the electric-potential questions incorrectly also answered the rate-of-change questions incorrectly. They then suggested that curricula designers should take that correlation into account and approach the concept of electric potential by first reviewing the concept of rate of change and then bridge it to the concept of electric potential. They also suggested that instructors first use easier rate-of-change situations, such as gradients of a hill, and graduate toward harder rate-of-change situations, such as electric potentials.

Still other scholars have been interested in looking at how students approach problem solving of electric fields. Greca and Moreira [14] observed college students' performance in physics classrooms and on labs, homework, and exams to see whether students construct comprehensible mental models when they solve problems. They found that some students in the study constructed mental models, whereas other students did not construct mental models but worked mostly with propositions—definitions, formulas, etc.—and manipulated them routinely to solve problems. Besides, they pointed out that there were many students who constructed incomplete mental models and understood the concepts only partially. Identifying this range, Greca and Moreira indicated that we should not treat a large group of students as an undifferentiated class of novices, but should recognize the substantial differences among them.

Furio and Guisasola [2] administered a questionnaire to investigate high school and college students' ontological and epistemological beliefs about electromagnetic interactions. They reviewed the historical development of theories of electromagnetism with respect to the evolving epistemology held by scientists in different eras. They then compared the historical epistemologies with the epistemologies identified from students' responses. Furio and Guisasola suggested that students usually understood electric interactions as acting at a distance: two charges separated in space can exert forces on each other instantly with no necessary medium in between. This idea, according to Furio and Guisasola, was undergirded by the Newtonian cosmology, which indicates that "forces are considered particular aspects of material interactions without procedural explanation" (p. 515). The action-at-a-distance conception of forces prevailed when Coulomb generalized the law of electric interaction in the late 1700s and the idea of a field was hardly formulated. After Faraday and Maxwell constructed the field theory of electromagnetism in the 1800s, electric interactions were explained as contiguous forces, transmitting in a limited speed through a medium—the electric field—between charges. This field point of view is grounded in what Furio and Guisasola called Cartesian cosmology: "forces happen by means of the vortexes or whirls that emanate from corporeous matter" (p. 515). The idea of a field was demonstrated as more accurate than the idea of action in distance in explaining electromagnetic forces and became the canonical theory of electromagnetism. Furio and Guisasola [2] argued that most students adopted the Newtonian cosmology rather than the Cartesian cosmology when solving problems about electromagnetism; students saw an electric interaction as an action at a distance instead of a contiguous force and therefore continued to experience difficulties understanding the concept of electric fields.

The aforementioned studies explored students' understandings of electric fields with specific conceptual foci. Other researchers have been trying to assess students' understandings of electromagnetism more broadly and have identified a collection of student misconceptions. By administering multiple questionnaires, Sağlam and Millar [15] found that high school students confused electric fields with magnetic fields; students saw field lines as indicating a flow from positive charges to negative charges just like electric currents flow from the anode to the cathode in an electric circuit; students used a cause-effect reasoning in situations where it does not apply (similar to what was found by Viennot and Rainson in 1992); and students had trouble dealing with effects associated with the concept of rate of change (similar to what Allain and Beichner found in 2004).

Maloney and colleagues [1], employing questionnaires, surveyed students' comprehension before and after a college physics course of electromagnetism. They found that some ideas were prevalent among students and difficult to change even after the course. For example, students did not explain the action and reaction between two charges as equal and opposite; instead, student responses indicated that they thought the larger charge exerts a greater force. Other findings that resonate with research by Viennot and Rainson [4] and that by Sağlam and Millar [15] include the following: students were confused by the superposition of fields generated by multiple charges; students mistook magnetic field effects for electric field effects; and before instruction, students were apt to make analogies between electric field lines and electric currents. Maloney et al. [1] have also called for further investigation into where these ideas come from and whether they are rooted in common sense much in the same way that students studying mechanics tend to think *force implies motion*.

## III. RESEARCH ON STUDENT LEARNING OF REPRESENTATIONS OF ELECTRIC FIELDS

Some of the studies reviewed above have included the canonical representation of electric fields, electric field lines, as part of their surveys and have tested students' interpretations [1,15]. There are also studies that have been devoted to studying student conceptions of electric field lines. Tornkvist and colleagues [3] gave college students a paper-and-pencil test with questions about electric field lines as the students were taking an electromagnetism course. The questions asked students to find the conceptually mistaken lines (the crossing lines, the sharply bent lines, and the closed lines) in a diagram consisting of charged conductors and corresponding electric field lines in that space (see Fig. 1, left). They also interviewed students, exploring their understandings of electric field lines. During the interview, students were presented with printed diagrams of canonical field lines. The interview questions centered on the concept of electric field strength, electric forces, and the charge's velocity and trajectory in each of the situations depicted in the diagram. (For example, students were asked to show the force on and the trajectory of a test charge at a given point in a given field represented by electric field lines, Fig. 1, right.) Based on an analysis of students' responses, they concluded that students did not grasp the relation between the following concepts: charge position, field line, force vector, velocity vector, and trajectory of a test charge. As a result, students confused field lines with other types of representations, such as force vectors, velocity vectors, and trajectories.

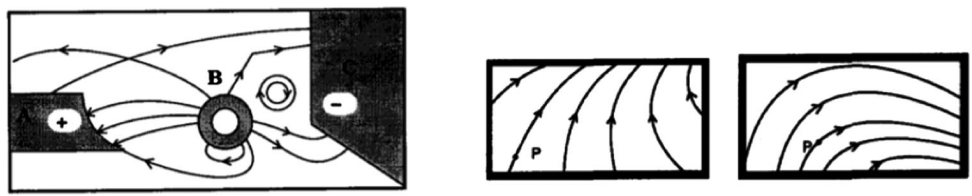

FIG. 1. Electric field lines presented to students in Tornkvist *et al.*'s study (1993). Left: The error finding task. Right: Two examples of the tasks administered during the interview.

Visualizing field lines by way of technologies has been tried to help students learn electric fields. Belcher and Olbert [16] developed 3D animations of electromagnetic field lines and suggested that the animation could reinforce users' insights into the connection between the shape of field lines and the dynamic effects of electromagnetic fields. Shortly thereafter, Dori and Belcher [17] conducted a long-term study investigating undergraduate students' comprehension of electromagnetism when they took an innovative course that adopted the 3D animation [16] as an innovation (others included small group discussion with intensive experimentation replacing lectures). To measure the effects of the new curriculum, Dori and Belcher [17] administered pre- and postassessments, which consisted of conceptual questions from standardized tests. Their results showed that the group taking the innovative course had greater conceptual gains than the control group taking a more traditional course.

A computer simulation, the Electric Field Hockey, developed by PhET (a University of Colorado educational technology group that designs interactive simulations for educational purposes) is another example of visualizing electric field lines through computer technology [18]. Electric Field Hockey simulates electric forces in a scenario in which electrically charged balls are pulling and pushing an electrically charged hockey puck. A learner who interacts with this simulation can manipulate the charges and the forces to determine the hockey puck's trajectory and destination (to hit the goal). The simulation makes multiple representations (field lines, force vectors, trajectories) visible to the learner to make sense of the physical concepts in relation to electric fields. To date, no empirical study has been conducted about the use of the Electric Field Hockey simulation.

## IV. STUDENT IDEAS AS FLEXIBLE LEARNING RESOURCES

Piecing together the research reviewed in the previous section into a larger mosaic, we can see that so far researchers have explored students' ideas of electric fields; accomplished their research objectives by looking at students' answers to textbook-style questions; surveyed students while they were learning about electric fields; primarily studied college students; and mostly focused on students' fixed and inaccurate understandings of electric fields. When studying representations, past researchers presented to students the canonical representations of electric fields and focused on students' interpretations. This mosaic reveals gaps. Currently, the literature has limited studies with *high school* students; it lacks knowledge about students' *preinstruction* ideas and ideas expressed in *open-ended tasks*; little work has been done addressing the *flexibility and variability* of student ideas; little is known about *student-produced* representations. These gaps limit researchers' ability to find out about students' productive ideas and to improve instruction from a constructivist perspective.

Taking a constructivist standpoint, we view learning as building new understandings upon previous ones and conceptualize students' ideas, canonical or noncanonical, as learning resources [7,8]. According to Hammer and colleagues [7,8] certain cues lead learners to activate prior ideas and assemble the ideas to make sense of the task at hand. The ideas activated are often fragmented, loosely connected, sensitive to context, and grounded in previous experiences. These prior ideas are precious resources on which students can build more systematic, coherent understandings. In the present study, we hoped to elicit high school students' ideas of electric fields preinstruction and analyze them from a resources perspective. We deliberately targeted the resources that connect to students' life experiences, for, according to Hammer and colleagues [8], these resources more often lead students to a sense-making process than does textbook formalism.

Besides the learning resources framework [7,8], another closely related theoretical framing we use in this study is related to student-produced representations. We perceive student-produced representations (including drawings, diagrams, and others) as artifacts that reflect students' ideas. Previous studies have shown that different aspects of student thinking can be exhibited when students are producing representations of a given topic. For example, diSessa and colleagues [5] identified sixth-grade students' conceptual and social skills exhibited in re-inventing Cartesian graphs when the students were representing a moving object. Sherin [19] identified "constructive resources" (p. 401) that can contribute to students' invention of novel representations of motion. The constructive resources identified in Sherin's study [19] include drawings that had been acquired through students' previous experiences. Studies about primary students on student-created visual models [6] and drawings [20,21] have also shown that visual models and drawings can serve as tools for students to comprehend related concepts or to develop scientific reasoning abilities when the targeted topics are often deemed challenging at that grade level. Inspired by past research, in the present study we specifically prompt students to produce drawings for scenarios of electric interactions, hoping to tap into students' ideas through these artifacts.

In sum, in the present study we explore (i) what preinstruction ideas of electric fields high school students can have and (ii) how the ideas emerge and vary in different contexts.

## V. METHOD

### A. Participants

Participants in this study were 92 ninth grade students enrolled in a summer program at a private, after-school educational organization in China. The summer program provided academic courses (English, math, science, etc.)

developed to help students transition from ninth to tenth grade (the cutoff between middle and high school in China). In these courses the teachers covered the first few units that would be officially taught in high school the following fall. At the time when the students participated in our study, most of them were between 15 and 16 years old and had not yet received formal instruction on electric fields. They had not systematically learned Newtonian mechanics but had learned about objects at rest experiencing two balanced forces.

The physics course in the summer program consisted of 12 lessons lasting 90 minutes each. The first author of this paper taught the physics course to three groups of students and implemented a lesson on electricity during the last session (lesson 12) with each group. In lessons one through ten, the teacher covered one-dimensional kinematics. In lesson 11, she briefly covered force composition.

The 92 students who participated in this study were in the three groups mentioned above: 33 students from group 1, 29 students from group 2, and 30 students from group 3. Group 1 was at the main site of the after-school organization. Groups 2 and 3 were at a satellite site in the same city. Most of the students lived in and went to school in the city where the organization was located. The students came from a variety of socioeconomic backgrounds.

### B. The lesson

The goals of the electricity lesson were (i) to expose students to phenomena of electric interaction and (ii) to give students an opportunity to produce representations of electric interaction. To achieve the first goal, we had students play with the computer simulation mentioned earlier, the Electric Field Hockey [18]. We used virtual electrostatic phenomena to substitute for the real-world phenomena [22] due to the unavailability of lab equipment in the place our study took place. To achieve the second goal, we used a worksheet asking students to draw comic strips about electric charges as if they were cartoon characters. Our focus in this paper is a study of students' written work in the comic strip activity. What happened during the Electric Field Hockey activity helped us to understand the context in which the comic strip activity was introduced.

#### 1. The electric field hockey activity

At the beginning of the electricity lesson, the teacher announced to the students that in the following activities there were no right or wrong answers and the main goal was for students to share ideas. She also informed the students that these activities were part of our research project. The Electric Field Hockey simulation was installed on the teacher's computer, which broadcast its screen through a projector onto the front wall of the classroom. One student at a time held a wireless mouse and operated the simulation, while the other students were watching and providing suggestions. The teacher helped to circulate the mouse in class, making sure that different students got a chance to try. The task of the simulation is to, in a virtual environment, move a hockey puck (representing a positive test charge) by creating pushes and pulls from multiple other charges and eventually to hit the goal (see Fig. 2). Source charges of the electric field are represented by red and blue balls for the different types, stored separately in the boxes in the top right corner, and can be drag and dropped in the hockey field. The simulation has four choices of settings: the practice setting has a blank field, and the three challenge settings with difficulties 1 through 3 have one to three vertical barriers standing in the middle of the field. In the challenge settings, the hockey puck must get around the barriers to hit the goal. During the electricity lesson, we had students play with two settings: the practice setting (see Fig. 2, left) and the challenge setting of difficulty 1 with one vertical barrier in the middle of the field (see Fig. 2, right).

Figure 3 shows the reproduced screenshots of the simulation that show students' strategies and results during the practice and difficulty 1 settings.

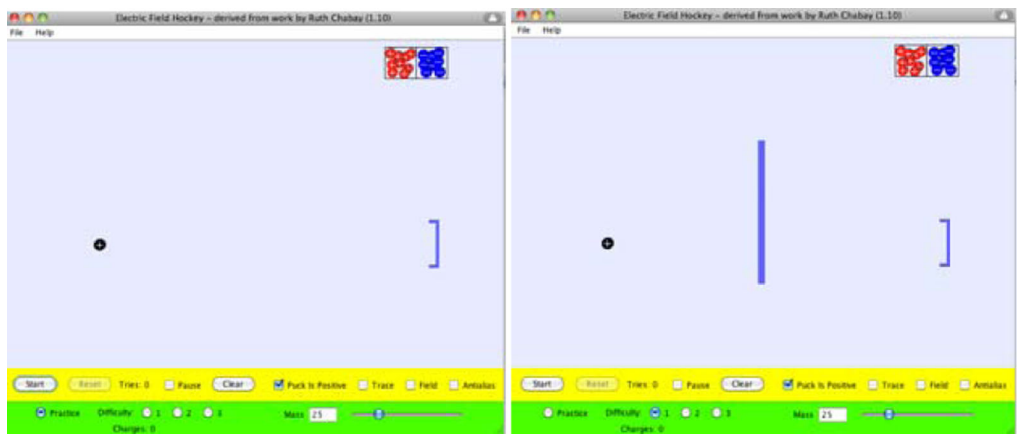

FIG. 2. Screenshots of the Electric Field Hockey game. Left: The Practice setting. Right: Difficulty 1 setting.

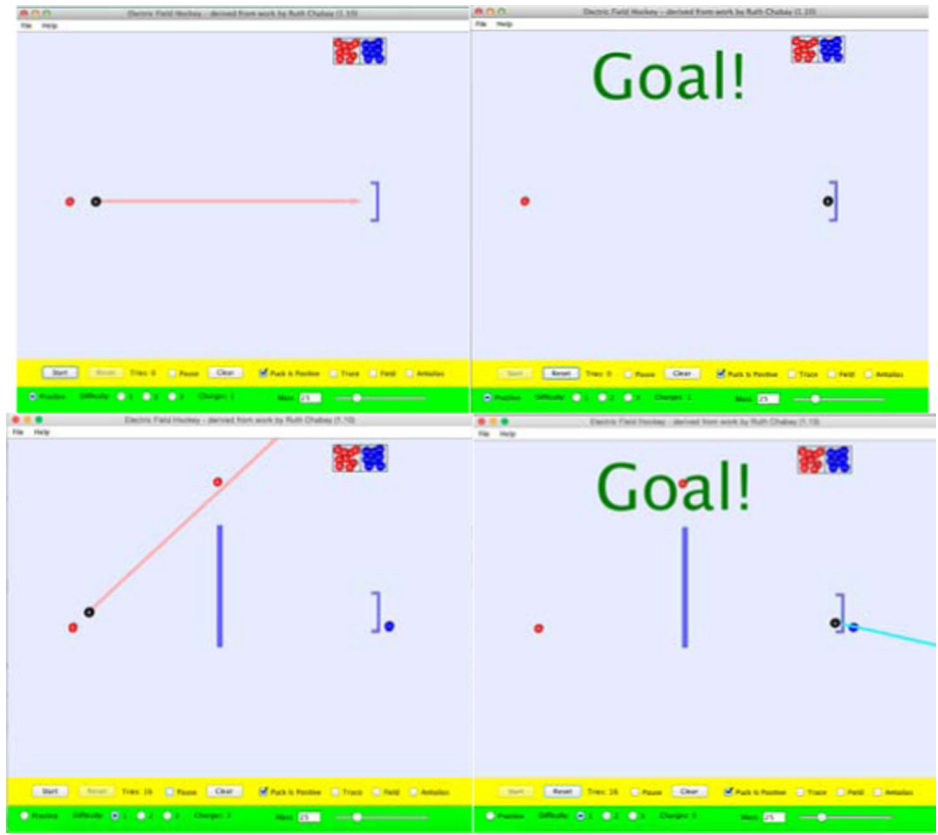

FIG. 3. Reproduced screenshots of students playing the Electric Field Hockey game. Top left: Students' strategy for the Practice Level. Top right: Result of students' strategy. Bottom left: Students' successful strategy for Level 1. Bottom right: Result of students' strategy.

In class, students easily completed the task of the practice setting at the beginning of the lesson.[1] They put a red ball to the left of the hockey puck that was at its starting position (see Fig. 3, top left) and hit the "start" icon. The hockey puck then moved to the right and successfully hit the goal (see Fig. 3, top right). As shown in Fig. 3, an arrow diagram automatically appeared right after the charged balls were dropped in the field, which represented the electric forces the hockey puck experienced. During the lesson, the teacher did not address these arrow diagrams and the students did not explicitly ask about them.

After accomplishing the task of the practice setting, students moved on with the difficulty 1 setting, which had a vertical barrier in the middle of the field between the hockey puck's starting point and its goal. Students tried for several times to arrange the charged balls in different patterns that they thought would make the puck get around the barrier and hit the goal. The puck instead always hit on the barrier and was stopped. Students spent more than 10 min on this setting, but still did not succeed. At 13 min into the lesson, the teacher concluded the Electric Field Hockey activity and started the comic strip activity, on which students worked until the teacher announced a short break.

During the break, a few students gathered around the teacher's computer and resumed the Electric Field Hockey simulation at difficulty 1. After some more tries, the small group of students finally made the hockey puck hit the goal and shouted out with cheers. The teacher and some other students in the classroom saw that happen. When the break was over and students came back, the teacher asked the students who played the simulation during the break to come up and share their strategy with the whole group. Guanhua (one of the three students we later interviewed) stepped up to the teacher's computer (with the screen projected on the board) and arranged the balls as shown in Fig. 3, bottom left, and clicked on the start button. The whole group watched the hockey puck moving around the barrier from the top and hit the goal. Students shouted hooray. The teacher asked the group to think about Guanhua's strategy and had students draw on paper strategies that they thought would succeed. Some students did this and some did not. About 10 min into the second half of the lesson, the teacher resumed the comic strip activity (see Table I).

---

[1]The description is drawn from group 1 that had the students we interviewed after class.

TABLE I. Lesson timeline.

| Activity | Time (min) |
|---|---|
| Electric Field Hockey simulation | 13 |
| Section 1 individual work | 5 |
| Section 1 whole group sharing | 5 |
| Section 2-1 and 2-2 individual work and whole group sharing[a] | 17 |
| Break | 10 |
| Electric Field Hockey simulation | 10 |
| Section 3 individual work | 4 |
| Section 3 whole group sharing | 4 |
| Section 4 individual work | 4 |
| Section 4 whole group sharing | 4 |
| Section 5 individual work | 4 |
| Section 5 whole group sharing | 5 |
| Wrap up | 5 |
| Total | 90 |

[a]The camera ran out of battery and stopped recording at one point of this session, so we do not have more specific times during this period.

### 2. The comic strip activity

The main activity of the electricity lesson was the comic strip activity: a worksheet of drawing comic strips about interactions between positive and negative charges as if they were characters in a cartoon series (the initial scene of each story was given). In the worksheet (summarized in Table II), we prompted students to represent electric interactions through visual stories and in their own language. We asked students to draw multiple frames for each story to enlarge the probability of representing the intermediate steps in the process of charge interactions. We also developed different sections in the worksheet intended to tackle students' ideas of electric forces (Secs., 1, 2-1, and 2-2) and of electric fields (Secs. 3–5).

As shown in Table II, Secs. 1, 2-1, and 2-2 concern charge interactions with the presence of a test charge and ask students to draw a four-frame story for each scenario. In Sec. 1, a positive charge sits in the room and is immobile when a negative charge enters the room. In Sec. 2-1, a big positive charge and a big negative charge both sit in a room and are unable to move when a little positive charge enters the room. In Sec. 2-2, two big negative charges sit in the room when a little positive charge enters the room. As we designed the worksheet, the charges that we described sitting in the room and immobile were intended to represent charges exerting forces, or source charges of the electric field. The charge that enters the room in each scenario was intended to represent the test charge that experiences forces and reflects the effects in its motion. To avoid both ourselves and students using physics terminology, we named the charges as if they were people: Little Positive, Little Negative, Big Positive, and Big Negative.

Sections 3, 4, and 5 made the test charge absent and asked students to show the influence the big charges can have on the test charge if it were present. Section 3 uses the two scenarios that were in Secs. 2-1 and 2-2, and Sec. 4 makes each room have one single positive or negative charge. Section 5 asks students to show the influence of a pair of parallel charged plates on a test charge in the area between them. Student worksheets were handed out in packets to each individual student. When drawing the comic strips, students started with Sec. 1 and worked on it individually. The teacher walked around and looked at the student's work. After most students finished the section, we called on a few students to project their work and share with the whole group. When choosing students, we purposely looked for diverse representations that she had noticed the student's work. The whole group then moved on to work with the next section in the same pattern. Table I shows the times spent in different sections each activity.

### C. Data collection

We videotaped the lesson with the three groups, collected 92 students' work, and interviewed individual students after class about their work. Videotaping allowed us to document the entire class and understand the context of each activity. Collecting and reviewing student work helped us notice the general range and patterns that emerged in student work to decide which students to interview. Interviewing was to ascertain in detail student ideas expressed in their work.

As we reviewed student work, we noted the variety in students' productions, such as drawings of vivid cartoon figures, delicate room decorations, detailed character conversations, and abstract symbols and diagrams. However, despite the diversity in students' productions, the arrow diagram was commonly included in students' work: of the 92 students, 90 drew the arrow diagram at some point in their work. On the other hand, there was still wide variation in the number of arrow diagrams individual students included in their work and the context in which individual students introduced the arrow diagram.

Arrow diagrams are the canonical graphic representation of electric fields. Previous studies [1,3,15] have suggested that students confused the arrow diagrams that represented field vectors with the arrow diagrams that represented other physical concepts such as velocity, trajectories, and electric currents. Students in our study spontaneously drew arrow diagrams even when the worksheet prompted them to draw cartoon comics. We were intrigued by this phenomenon and wanted to delve into further detail: What did the students mean by the arrow diagrams in their work? How did their understanding of electric fields compare to what the literature has suggested? Therefore, we decided to do in-depth interviews on students' use of arrow diagrams.

We chose students from group 1 to interview for scheduling convenience. Within this group we selected the students whose work had arrow diagrams through all five sections of the comic strips. After checking students' assents and their parents' consents, five of them had assented and consented to participate in the post-class interview. We were able to schedule interviews with three out of these five students: Xinmiao, Mandi, and Guanhua.

TABLE II.  Comic Strip Activity tasks.

| Text | First scene |
|---|---|
| Section 1 Draw Something: Draw a four-frame comic telling a story about two characters: Little Positive and Little Negative. Little Positive is sitting at the center of a room when Little Negative comes in through the door. What will happen next? Use four frames to draw at least four scenes of the story. The picture shown could be the first panel, but feel free to draw your own first panel. You are not required to draw every character vividly; the key is to show how and why your characters act the way they do. | 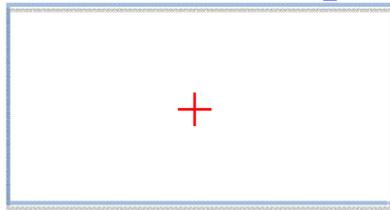 |
| Section 2-1 Draw Something: Draw a four-frame comic telling stories of two big charges and one little charge. In this first story, Big Positive and Big Negative sit in the room at a distance. They are fastened to their seats and cannot move. Little Positive comes into the room. What will happen next? Explain your story to the class. What do all the things that you have drawn mean? | 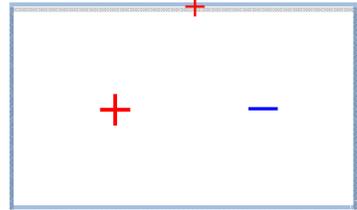 |
| Section 2-2 Draw Something: Draw a four-frame comic telling stories of two big charges and one little charge. In this second story, both big charges sitting in the room are negative (Big Negative and another Big Negative), while the little charge is positive (Little Positive). What will happen when Little Positive enters the room? Explain your story to the class. | 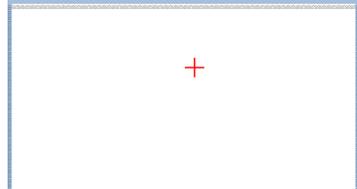 |
| Section 3 Draw Something: Now what if we take away Little Positive from the previous two pictures? Can you find a way to show the influence the big charges had on Little Positive that had been present in the previous two sections? | 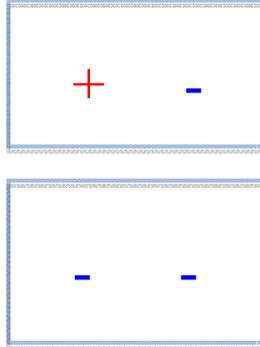 |
| Section 4 Draw Something: What if we leave only one big charge in the room, Big Positive or Big Negative? How can you show the influence the big charges have, respectively, on Little Positive that might come into the room? | 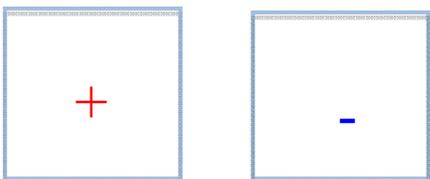 |
| Section 5 Based on what you have drawn before, how can you show the influence that the two charged plates have on Little Positive that moves between the plates? Why? | 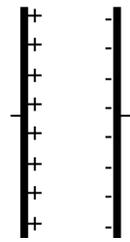 |

Y. C. carried out the post-class interviews. Xinmiao was interviewed 10 days after the date when the electricity lesson was taught. The interview lasted 54 min. Mandi was interviewed 11 days after the electricity lesson and the interview lasted 51 min. Guanhua's interview was conducted 16 days after the electricity lesson and lasted 40 min. Each interview was video recorded. During the interview, we asked the student to explain his or her work on each section, frame by frame. We also asked clarification questions about specific parts of their work within a frame. If the student changed his or her mind or came up with new ideas about a section, he or she was allowed to revise and asked to explain the new work.

### D. Data analysis

We did qualitative analysis on the three students' written work and the interview data and constructed three case studies [23], one for each student. In each case, we included a thick description [24] of the student's written work, his or her explanation, and our interpretation. The goal of the case studies was to understand these three students' ideas of electric fields in context.

As the first step of this process, we transcribed the videos of all three interviews (in Chinese) and translated them into English. To test the accuracy of the translation, we asked a graduate student who was a native Chinese and studying physics in a United States graduate school to read a randomly selected excerpt of the transcript while watching the corresponding video clip. There was no disagreement on the translation.

After transcribing and translating, we carried out a line-by-line examination of the transcript. We flagged the moments when students explained the meaning of an arrow diagram. We marked the lines when the student was describing an idea that had been discussed in the literature. At the same time, we annotated any emergent themes in the students' explanations. We first independently looked at a randomly selected excerpt of the transcript, and together we discussed the annotations until we reached an agreement about our notes. According to these notes, the first author read and annotated the rest of the transcripts and the second author read and provided comments on the annotations. Together we then discussed the transcripts, the annotations, and the comments until we agreed on our interpretation of the students' understandings.

In addition to the three individual case studies, we constructed a cross-case synthesis [23], highlighting the written work and oral explanations that reflected a particular conception of electric fields. Lastly, we compared our findings with the research claims made in previous studies in this area.

## VI. RESULTS

In the following, we will present Xinmiao's, Mandi's, and Guanhua's case studies. To address our two research questions concerning what student ideas are and how they vary, we will describe the ideas we identified and present in detail the situation in which each idea was generated and sometimes evolved.

### A. Xinmiao's case

Table III shows Xinmiao's work and explanation. The drawings in red ink were his original work produced in class. The black remarks were what he added during the interview: at times he wanted to highlight the drawings while talking; other times he reread the worksheet and realized that he had misunderstood the task and had not done it appropriately, so he redrew his work on the worksheet in black ink. His verbalized ideas are highlighted in italic.

Throughout all sections, Xinmiao expressed his idea that *opposite charges attract and like charges repel*. For example, in Sec. 1, he drew a positive sign and a negative sign in each frame and the distance between the two signs became closer in later frames. When explaining this piece of work, Xinmiao said,

> *"When the two charges met, they were attracted to and moved toward each other and combined into a new thing."*

In the third and fourth frames, the two signs were both drawn at the center of the frame. He said the fourth frame just repeated the third frame because the task required him to draw a four-frame comic strip. As shown in his explanation, he also referred to an idea that *two opposite charges ought to be combined*. This idea reoccurred in his later explanation for Secs. 2-1 and 2-2.

Xinmiao's explanation for Sec. 2-1 indicated that the arrows he had drawn in his work showed the *directions of the test charge's motion*. He said,

> *"The little positive charge came into the room and saw two charges. It looked for a big charge to combine with. It first went to the big positive charge but could not combine with it. It then moved to the big negative charge and combined with it."*

Xinmiao's work in Sec. 2-1 described the test charge interacting with multiple charges *sequentially*. The first frame depicted the moment before the little charge came into the room. The second frame showed that the little charge went to the big positive charge (but got rejected, according to his explanation). In the third frame, the little positive charge was shown moving to the big negative charge. In the fourth frame, the little positive charge combined with the big negative charge, which was the end of the story. After Xinmiao gave his explanation, the interviewer followed up and asked, "Why did you make the little positive charge go to the big positive charge first (instead of going to the big negative charge first)?" The following conversation took place.

TABLE III. Xinmiao's work.

| Section | Student work | Explanation |
| --- | --- | --- |
| 1 | 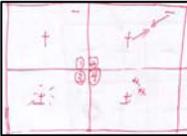 | Xinmiao: "When the two charges met, they were *attracted to and moved toward each other and combined* into a new thing." |
| 2-1 | 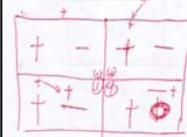 | Xinmiao: "The little positive charge came into the room and saw two charges. It looked for a big charge to *combine with*. It first *went to* the big positive charge but *could not combine with it*. It then moved to the big negative charge and *combined with it*." |
| Section 2-1 new | 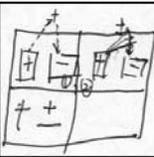 | Xinmiao: "When it (the little positive test charge) comes in, it sees that the big positive charge is repulsive, and it can't go there. So it goes directly to the negative charge." |
| 2-2 | 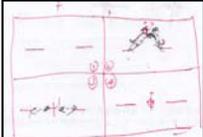 | Xinmiao: "The little positive charge saw both big negative charges *attracting it*, and it could *combine with* either of them. It did not know where to go. The two big negative charges' *attracting forces* were the same and thus *balanced*, so the little positive charge stayed in the middle." |
| 3 | 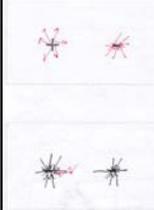 | Xinmiao said that his drawing meant that *a big positive charge could repel a little positive charge, a big negative charge could attract a little positive charge*, and this impact on the little positive charge was in all directions. |
| 4 | 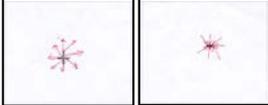 | Xinmiao claimed that this section was the same as Sec. 3. The only difference was that in Sec. 3, the big charges also *attracted or repelled* each other because they were in the same room. |
| 5 | 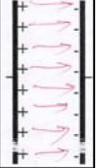 | Xinmiao said that the lines meant that the positive charges on the left plate were *attracted to* the negative charges on the right plate. In class, Xinmiao was called by the teacher to share this piece of work with the whole group, but he only said that he was not able to articulate the meanine of the red parallel arrow lines he had drawn. |

*Xinmiao: If the little charge went to the big negative charge first, they would combine directly; the story would just end.*
*Interviewer: What's wrong with that?*

*Xinmiao: That way I didn't get to talk about the interaction between the little positive charge and big positive charge; the story was incomplete.*
*Interviewer: What do you mean by "incomplete"?*
*Xinmiao: I wouldn't be able to fill up all four frames.*

Seeing that Xinmiao was concerned with filling in all four frames, the interviewer asked him to redraw this section without worrying about filling in frames. Xinmiao then drew a new piece of work, shown in the row labeled "Sec. 2-1 new." In his new work for Sec. 2-1, Xinmiao drew forces from multiple charges in the same frame, representing that *electric forces act simultaneously when multiple charges are present*. He also did this in his work for Sec. 2-2.

This clip of the interview showed that Xinmiao's representation of the interaction between multiple charges was affected by multiple local factors: his understanding of the interaction, his understanding of the task requirement, and the interviewer's request he received.

When Xinmiao explained his new drawing for Sec. 2-1, the interviewer asked him another question: "How come the little positive charge knew that it would be rejected or accepted by the big charges?" Xinmiao struggled for a while and answered, "because the big charges were sending out signals to the test charge." This understanding was close to the idea that *electric interactions transmit through space*. This understanding of "sending signals" seemed to be a common sense for him and could be a productive resource when he learns about electromagnetism.

Xinmiao produced multiple direction arrow diagrams in Secs. 3 through 5. Xinmiao said that his drawing meant that a big positive charge could repel a little positive charge, a big negative charge could attract a little positive charge, and this impact on the little positive charge was in all directions. According to his explanation, the arrow diagrams meant that *a charge can exert forces in any direction depending on where the little charge was*. The arrow diagrams in Xinmiao's work represented multiple meanings: *the direction of a force* and *the direction of a charge's motion*. He did not seem to be bothered by using the same representation to describe multiple meanings.

### B. Mandi's case

Mandi's work is shown in Table IV. Mandi repeated the idea that *opposite charges attract and like charges repel* through all the sections (similar to Xinmiao). Different from Xinmiao, who often brought up the idea of the combination of two opposite charges, Mandi did not mention the final state (combination or otherwise) of the charges at all. For example, Mandi explained her work of Sec. 1,

> *"The two charges attract each other and move toward the middle."*

It seemed that Mandi focused on the mechanism (attraction and repulsion) whereas Xinmiao focused on the mechanism and the result (opposite charges combine). Mandi's explanation in Sec. 2-1 also showed an idea that *a bigger charge exerts a stronger force*.

> *"[A] little charge may not attract as strongly as a big one, or… the strength of attraction [of a big charge] is bigger. … [W]hen (the little positive charge) comes in, it is immediately repelled by the big positive charge and attracted to the big negative charge. So the moving trajectory should be…"* (gestures a curve from left to right).

The arrow diagrams in Mandi's work in Secs. 3 through 5 have *lines always starting from positive charges and ending at negative charges*. In Sec. 3, she drew bundles of curved arrows that looked like the canonical electric field lines, although not accurate. She explained the shape and meaning of these lines with reference to what she had learned about magnetic field lines. She also brought up the phrase "electric field lines" when she explained.

> *"Because magnetic field lines start at the North pole and end at the South pole, electric field lines should start from the positive charge and end at the negative charge."*

Mandi seemed to have a lot of ideas about these lines. She always drew charges at both ends of a field line: none of the line had a loose end. When there was only one charge provided in the frame, she added in a little charge for the lines to stop or start (as in Sec. 4). The shapes of the lines were all curved. She said that *those lines must be curved lines so that they could fill up the space*. Mandi said that *the electric field lines should be like magnetic field lines* because *electricity and magnetism are closely related*. Mandi also claimed that those lines were *not real lines*. When Mandi said that the lines were *not real*, we were not sure what she meant by "not real." The interviewer asked Mandi to explain herself, but Mandi just repeated that they were "not real." It sounds like she was repeating a statement that she had previously learned. She did say a couple of times that all these ideas came from what she had learned in middle school.

The meaning of these arrow lines, according to Mandi, were the attraction and repulsion from the big charges to the little positive charge. She said that the curved lines also represented the trajectory of the little charge because she had seen in the Electric Field Hockey simulation that the hockey puck moved along a curve.

Some of Mandi's drawings looked like the canonical field lines of the electric field of multiple charges. Although she might not yet understand the canonical meaning of electric field lines (i.e., collections of numerous local electric field vector), she was able to draw on her ideas of magnetic field lines and create electric field lines that made sense to her. We view this as a possibly productive connection as she had something to help her understand a new topic that she had not seen before.

### C. Guanhua's case

Guanhua was the last student we interviewed. During the interview, his ideas evolved and he added new work to multiple sections except for Sec. 1 (see Table V).

TABLE IV. Mandi's work.

| Section | Student work | Explanations |
|---|---|---|
| 1 | 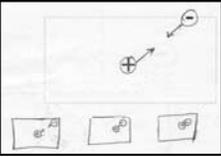 | Mandi: "The two charges *attract each other and move to the middle*." |
| 2-1 | 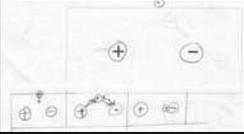 | Mandi: "[A] little charge may not attract as strongly as a big one, or… *the strength of attraction [of a big charge] is bigger.* … [W]hen (the little positive charge) comes in, it is *immediately repelled by the big positive charge and attracted to the big negative charge*. So the *moving trajectory* should be…" (gestures a curve from left to right). |
| 2-2 | 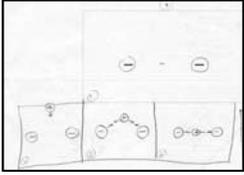 | Mandi: "Because the strength of the two negative charges was the same, their attraction to the little positive charge should also be the same. [.] The little positive charge moved to the middle point, where the force was the smallest, and got stuck there." |
| 3 | 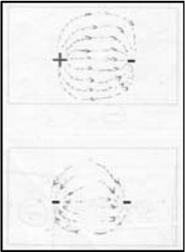 | Mandi: "Because *magnetic field lines* start at the North pole and end at the South pole, electric field lines should *start from the positive charge and end at the negative charge*." "The *lines show the attraction or repulsion to an imagined little positive charge in the middle of the two big charges*." "They are curves so they can fill up the space." "The little charge moves along these curve lines." |
| 4 | 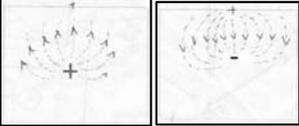 | Mandi: "The lines show the *attraction or repulsion from the big charge at the center to the little positive charge* at the top." |
| 5 | 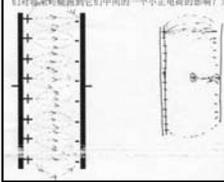 | Mandi: "The little positive charge in the middle *was repelled by the positive charges* on the left plate *and attracted to the negative charges* on the right plate." |

During the interview, Guanhua kept saying *opposite charges attract and like charges repel* to explain his work, as did Xinmiao and Mandi. Additionally, Guanhua seemed to have a model to explain the stories in Secs. 1, 2-1, and 2-2: the little charge came in with an initial direction (indicated by a little arrow attached to the little charge); at that moment, *the little charge was far away from the big charges in the room, the attraction and repulsion were weak*, so the little charge moved according to the initial direction. When the charges lined up in the same horizontal line, *they were the closest and the attraction and repulsion were the strongest*. From that point, the little charge's movement was determined by the directions of attraction and repulsion; the little charge moved accordingly and reached a final state.

In his explanation, Guanhua brought up the idea that *electric interactions are stronger when the charges are closer*. He continued to exhibit this idea in his work for Secs. 3 and 4, in which he drew circles around the big charges, denoting that *the electric interactions have a range of effect*.

An idea that was expressed throughout Guanhua's explanation, but not in Xinmiao's or Mandi's, was to represent the forces and movements in *orthogonal directions* (the "right directions," in his words) and separate these directions from the diagonal directions. For example, in Secs. 1, 2-1, and 2-2, Guanhua set initial velocity of the little charge in the vertical direction. He then drew all the charges lined up horizontally, in which case the forces and the moving directions were all in the horizontal direction. Only after the interviewer probed him about why it had to be in these directions did he start to talk about forces in the diagonal directions and add them to his work (see Table V, Secs. 2-1 and 2-2). Similarly, in Secs. 3 and 4, Guanhua kept bringing up the difference between the situation in which the test charge came in with a velocity in an orthogonal direction (drawing straight trajectories in these situations) and the situation in which the test charge came in with a velocity that was not in an orthogonal direction (drawing curved trajectories in these situations).

TABLE V.  Guanhua's work.

| Section | Student work | Explanations |
|---|---|---|
| 1 | | Guanhua: "The big positive charge said, 'Come on.' The little positive charge came in with an *initial moving tendency* pointing down, saying: 'I'm coming.' When the two charges were *in the same [horizontal] line, they were closest, so the attraction between them was the strongest.* The big positive charge wondered, 'Oh?' They *attracted each other and moved closer and got together*, saying, 'We are together!'" |
| 2-1 | | Guanhua: "The little positive charge came in with an *initial moving tendency* pointing down. The big positive charge said: 'We can't move.' When the three charges were *in the same [horizontal] line, the attraction and repulsion between them were the strongest. The big positive charge repelled the little positive charge, and big negative charge attracted it*. The little charge said, 'Keep moving.' At last, it came to the big negative charge and *combined with it*, saying, 'We are together!'" |
| 2-2 | | Guanhua: "The little positive charge came in and moved down. When the charges were *in the same straight [horizontal] line, the two attracting forces were the strongest*, and the *two forces were in balance*. The little positive charge was *stuck in the middle*." |
| 3 | | Guanhua: "The arrows heads in the middle of the lines *represented force directions: big positive charges repelled [a little positive charge] outward and big negative charges attracted [a little positive charge] inward*." |

TABLE V. *(Continued)*

| Section | Student work | Explanations |
|---|---|---|
| 4 | 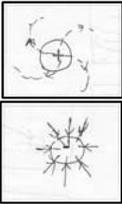 | Guanhua's explanation for Sec. 4 was the same as for Sec. 3. |
| 5 (Original) | 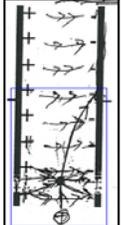 | This includes Guanhua's new work (produced during the interview) on top of his initial work in class. Rows below are the reproduction of his work, layer-by-layer, reflecting his thinking process during the interview. |
| 5 (Reproduced, layer 1, in class) | 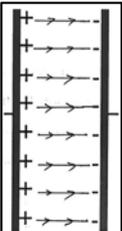 | Guanhua: "The parallel lines with arrows pointing to the right meant *the positive charges on the leftplate were attracted to the negative charges on the right plate*." |
| 5 (Reproduced, layer 2, interview) | 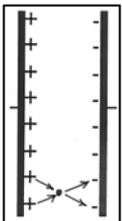 | Guanhua: "When there is a little positive charge in the area between the plates, *it is repelled by the nearby positive charges and attracted to the nearby negative charges. The forces were in opposite directions so they should be in balance and the little positive charge should stay at rest*." |
| 5 (Reproduced, layer 3, interview) | 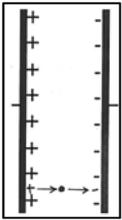 | Guanhua: "Because you don't limit the number of charges," he said, "I can add one more pair. *The little positive charge is repelled by the positive charge [just added] and attracted by the negative charge [just added], so it will move*." |
| 5 (Reproduced, layer 4, interview) | 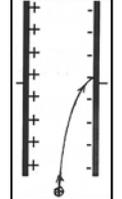 | Guanhua: "It will *keep being pushed to the right along the way after it entered the area between the two plates. As a result, it will move along a curve (draws the parabola) until it hits the right plate* and then *continue to move upward and be pushed tightly against the plate*." |

Guanhua developed more sophisticated ideas of the superposition of electric forces while he was explaining his work during the interview. For sections that had multiple charges exert forces on the little test charge, Guanhua drew the forces in the same frame and showed the idea that *forces act simultaneously*. By the end of the interview, he demonstrated his understanding of the *composition of forces from multiple charges* and used that to *determine how a test charge would move accordingly*.

Another idea Guanhua expressed throughout the interview was about *initial velocity* (the "moving tendency," in his language). He attributed initial velocity to the little charge in Secs. 1, 2-1, and 2-2 and talked about how the velocity determined the little charge motion in the beginning of each scenario. In Secs. 3 and 4, Guanhua drew curved arrow lines around the big charges because, as he explained, *when the initial velocity of the incoming little positive charge was not in an orthogonal direction, the little positive charge's trajectory would be a curve*. He had drawn straight trajectories around the big negative charges. After explaining his idea to the interviewer and rethinking it, he decided to change the straight ones also into curves.

Guanhua initially drew parallel arrow lines in Sec. 5 (Table V, Sec. 5, reproduced layer 1) to represent "the positive charges on the left plate were attracted to the negative charges on the right plate." When explaining Sec. 5 during the interview, Guanhua read the task text again and realized that it asked to represent the influence of the plates on a little positive charge. He realized that he had misunderstood the task and started to add new drawings on top of his old drawing. He first included a little dot and drew four diagonal arrows around it (Table V, Sec. 5, reproduced layer 2) and explained,

"When there is a little positive charge in the area between the plates, it is repelled by the nearby positive charges and attracted to the nearby negative charges. The forces were in opposite directions so they should be in balance and the little positive charge should stay at rest."

According to his gestures, a force pointing upper right and a force pointing lower right were what he meant by "in opposite directions."

He then added a pair of opposite charges on each plate that were in the same horizontal line with the little positive charge and drew the two arrow lines (Table V, Sec. 5, reproduced layer 3) and said,

"Because you don't limit the number of charges, I can add one more pair. The little positive charge is repelled by the positive charge [just added] and attracted by the negative charge [just added], so it will move."

He started to become confused by the two contradicting conclusions he had arrived at—the little positive charge will stay at rest or move. Then he recalled what he had learned about the principle of force composition and realized that in the situation of the previous row (reproduced layer 2), the "net force" (he later called it "the total pushing force") should be nonzero and to the right. He concluded that the little positive charge would definitely move to the right.

After he resolved the motion of a little charge that was inside of the parallel plates, he added a little positive charge that started outside of the plates and flies in from the bottom (Table V, Sec. 5, reproduced layer 4) with an upward initial velocity ("moving tendency," in his language). He drew a parabola to represent the trajectory of the little charge and explained:

"It will keep being pushed to the right along the way after it entered the area between the two plates. As a result, it will move along a curve (draws a parabola) until it hits the right plate and then continues to move upward and push tightly against the plate."

He figured out the trajectory of a little charge, with and without initial velocity, between the two parallel plates with opposite electric charges, and for the first time called the horizontal arrow lines he had drawn between the plates *force lines*. He summarized in the end:

"When there is no initial velocity, the little charge's trajectory is a straight line and same with the force line; if there is initial velocity, the little charge's trajectory is curved and different from the force lines."

This statement is fairly accurate in the situation he was presented with, and he arrived at this understanding by himself.

### D. Summary of the three students' ideas

In Table VI we summarize the ideas that Xinmiao, Mandi, and Guanhua exhibited in their written work and oral explanations. We summarize nine ideas that we identified and think are constructive to develop disciplinary knowledge of electric fields. Three of them (ideas 1–3) have not been the research foci in the literature of physics education research. Six ideas (ideas 4–9) relate to the literature—while prior studies have asserted that students had difficulties understanding them, we provide cases in which even younger and less exposed students

TABLE VI. Summary of students' ideas.

| Students' ideas | Evidence from students' work and explanations | Relation to literature |
| --- | --- | --- |
| 1. Opposite charges attract and like charges repel. | Mandi: "[W]hen [the little positive charge] comes in, it is immediately repulsed by the big positive charge and attracted to the big negative charge." Similar statements in all three students' explanations. | Not a focus in the literature. |
| 2. Opposite charges ought to be combined together. | Xinmiao: "The little positive charge came into the room and saw two charges. It looked for a big charge to combine with." Guanhua: "We are together!" | Not a focus in the literature. |
| 3. The attraction or repulsion from a source charge could be in any direction around it depending on the location of the test charge. | All three students' work (left: Xinmiao; middle: Mandi; right: Guanliua). 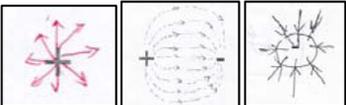 | Not a focus in the literature. |
| 4. Strength of electric force depends on the charges' amount of and the distance between charges. | Mandi: "[A] little charge may not attract as strongly as a big one, or… the strength of attraction [of a big charge] is bigger." Guanhua: "They were the closest, so the attraction between them was the strongest." | College students in previous studies failed to grasp the relationship between force strength and magnitude of charges and their distance. |
| 5. Electric forces act simultaneously when multiple charges are present. | All three students drew two force arrows from different charges in the same frame (left: Xinmiao; middle: Mandi; right: Guanliua). 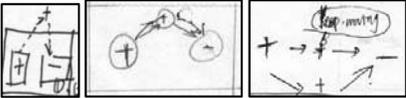 | High school and college students in previous studies had difficulties understanding the superposition of electric forces. |
| 6. Arrow diagrams have mixed meanings that combine both the direction of the force and of the charge's motion. | All three students said something related to this idea: They [the arrow lines] represent the forces as well as the [charge's] trajectory. | Similar to what was reported in previous studies with college students, the students in this study used arrow diagrams to depict force vectors and the trajectories of charges. |
| 7. Force lines and trajectories are different but related representations. | Guanhua: "When there is no initial velocity, the little charge's trajectory is a straight line and the same as the force line; if there is an initial velocity, the little charge's trajectory is curved and different from the force lines." | College students in previous studies were confused by these representations. |
| 8. Arrow lines are like magnetic field lines. | Mandi: "According to the fact that magnetic field lines start at the North pole and end at the South pole, these electric field lines should start from the positive charge and end at the negative charge. They should be curves and fill out the space." | Similar to what was reported in previous studies with high school and college students, students in this study related electricity and magnetism. |
| 9. Electric interactions are contiguous interactions, permeating space. | Xinmiao: "The big charges are sending out signals to the little charge [to let it know whether it is going to be attracted or repelled]." Mandi: "They [the lines] are curves so they can fill up the space." | High school and college students in previous studies understood the electric force as an action at a distance, not as a contiguous force. |

spontaneously leveraged these ideas. We will discuss these ideas more in the next section.

## VII. DISCUSSION

Following each subtopic in this section, we will discuss Xinmiao's, Mandi's, and Guanhua's ideas in relation to what has been reported in the literature. Xinmiao, Mandi, and Guanhua in our study were younger and less exposed to the topic of electric fields than most of the students in previous studies. The ideas they leveraged were nonetheless rich. Ideas 1–3 reflect the basic principles of electric interactions and have not been a focus of previous studies in physics education research. We speculate this is because previous studies have focused on students' learning difficulties and high school and college students do not have difficulties understanding these principles. Ideas 4–9 are about features and representations of electric fields that are more sophisticated than ideas 1–3. These ideas have been the research foci of previous studies reviewed earlier in this paper. We will next elaborate on the connections between our findings and the literature.

### A. Magnitude of electric forces

Students in our study provided evidence of their understandings of the magnitude of electric forces. Guanhua stated that *when the charges are the closest, the forces are the strongest*, which demonstrated his understanding of the relationship between the magnitude of the force strength and the distance between the interacting charges. Mandi also had the idea that *a bigger charge exerts a stronger force* (see Table VI, idea 4), which is an idea about the relationship between the force strength and the change amount.

According to Maloney *et al.* [1], many college students did not answer survey questions correctly about the magnitude of electric forces in relation to charge amount and the distance between charges. Our study shows that students can spontaneously leverage some aspects of a meaningful understanding, as in Guanhua's case. In Mandi's case, her statement about the magnitude of electric forces can be right or wrong depending on what specific forces she is referring to. If she was talking about two charges acting on a third charge (which, in the context of the worksheet, was legitimate because there were multiple charges in the scenario), the force strength does depend on the amount of electric charges and thus her idea made perfect sense. If, however, she was talking about forces between two interacting charges, then the forces should be equal in size and Mandi's statement was not accurate (as suggested in Maloney *et al.*'s study [1]). We did not follow up on this statement during the interview, and we think it is worth further exploration.

### B. Superposition of electric forces

When dealing with situations in which multiple electric forces were present, Xinmiao, Mandi, and Guanhua all drew the forces from multiple charges in the same frame, indicating that the forces were acting simultaneously (see Table VI, idea 5). In addition, Guanhua arrived at a fairly sophisticated understanding of the superposition of electric forces and of its impact on a test charge (see Table V, Sec. 5). In his initial work in class, Xinmiao did describe events in which the test charge interacted with multiple charges sequentially (see Table III, Sec. 2-1). He did so in order to make sure that he had addressed all the interactions and filled up all four frames that the worksheet seemed to ask for. Viennot and Rainson [4] have argued that students are inclined to apply a sequential reasoning to situations in which multiple factors are affecting simultaneously. Our study provides evidence of a student understanding that the factors acted simultaneously but representing them sequentially for other reasons, such as to respond to a task prompt and fill in a certain number of frames. Our data show that it is important to note local dynamics of student ideas. We saw Viennot and Rainson's [4] finding in Xinmiao, but it was specific to that context and it does not mean that Xinmiao did not understand the physics in question.

### C. Electric field lines

All three students spontaneously produced arrow diagrams that resembled canonical electric field lines (see Table VI, idea 3). The meaning that students attributed to these lines included force directions, velocity directions, and trajectories. Students were able to clearly articulate their explanations for the meaning of each diagram they had drawn (see Table VI, idea 6). This finding speaks to Tornkvist and colleagues' [3] claim that college students confused representations of electric fields and did not understand the relationship among the related concepts. The three students in our study were able to distinguish force from velocity verbally even though they represented the different concepts with the same symbol on paper. The students showed some understanding more complex than Tornkvist's interpretation [3].

Students did not necessarily express in-depth understandings of the representations in their first responses. Rather, more sophisticated understandings emerged while the students were explaining their work. For example, in his original work, Guanhua only drew the diagrams for special situations in which forces and motion were all in an orthogonal direction. As the interview progressed and he tried to explain his ideas to the interviewer, Guanhua started to talk about and added diagrams for the situations in which the forces were not in an orthogonal direction (the drawings he added to Secs. 2-1 and 2-2 during the interview). By the end of the interview when he explained his work for Sec. 5, he drew the force and motion diagrams correctly and analyzed the superposition of forces from multiple charges.

He related the net force to the test charge's initial velocity and figured out the test charge's trajectory. Ultimately, he generalized a sound theory about force, motion, and their representations, and completed a sophisticated diagram in his work (see Table V, Sec. 5).

By highlighting students' productive ideas, we do not mean to imply that high school students can do everything correctly without any instruction. Students in our study did show inaccurate understandings. For example, Mandi attributed double meanings—force and trajectory—to the curved lines she had drawn in Sec. 3 (see Table IV, Sec. 3). In these cases, force lines and trajectories cannot possibly be the same lines, so Mandi's understanding has flaws. Nevertheless, we would not broadly conclude that Mandi confused the representations. We argue for a more meticulous examination about the specific situation in which the representations were more likely to be misunderstood, for example, situations in which electric field lines are curved.

### D. Electricity and magnetism

Mandi brought up the connection between electricity and magnetism and addressed this connection frequently. She said that electricity and magnetism were closely related and in many aspects similar (see Table VI, idea 8). She drew field lines that started from the positive charge and ended at the negative charge. She did so not because she made an analogy between field lines and electric currents (as in Ref. [15]). Instead, she drew from what she had previously learned about magnetism (magnetic field lines start from the North pole and end at the South pole) and applied it to electricity. From a resources perspective [7,8], we see Mandi connect electricity to magnetism in an attempt to make sense of the topic by activating resources; the discrimination between electricity and magnetism can occur when students encounter scenarios in which recognizing the difference becomes necessary. Here we provide a different perspective than other researchers who have argued that students are apt to confuse magnetism with electricity [1,15].

### E. The concept of a field

Mandi brought up the words "electric fields" during the interview and said it should be similar and related to magnetic fields. Mandi also talked about the curved lines she had drawn, which she called electric field lines, and said that the lines should be curved because they need to fill up space. This statement reflects a primitive idea about the electric field: it permeates the space between electric charges (see Table VI, idea 9).

The other two students did not say the words "electric fields," but Xinmiao said that the source charges were sending signals to the test charge so the latter knew whether they were going to combine (see Table VI, idea 9). This statement reflects his thinking that the communication between electric charges needs a medium, which is a productive conception about electric fields, and a previous study [2] identified that college students did not often apply this thinking.

The comparisons we have made are to highlight what the three students can do rather than what they cannot. We hope to promote more research on understanding student thinking, examining student ideas in context, and recognizing variant learning resources [7,8].

## VIII. FOLLOW-UP WORK

We repeated this study the following summer with a different cohort of students, and collected a new set of data. In this second data set, we have analyzed more student work of the comic strip activity but focused on the idiosyncratic representations: cartoons students have drawn. We have also carried out an analysis on the Electric Field Hockey activity that preceded the comic strip activity. Papers reporting these analyses are in preparation.

Three years after we interviewed the three students in this study, we contacted them again in the summer to carry out postinstruction interviews. At that time, the students had just graduated from high school and should have learned electric fields in high school physics. We heard back from two students, Xinmiao and Guanhua. Both were going to college in the fall. We reinterviewed Xinmiao and Guanhua individually, presented them with the same worksheet of the comic strip activity, and asked them to draw their work. The interviews were recorded and student work was collected. We plan to carry out an analysis comparing the same student's pre- and postinstruction ideas of electric fields.

Because this study indicates that students' ideas are context sensitive, and the tasks used in this study were in many senses unique, one future direction of our work is to analyze the factors of a learning activity and examine how they could have shaped students' ideas, with a hope to make suggestions on activity design.

## ACKNOWLEDGMENTS

We would like to thank the summer school and the participant students for cooperating with us in this study.